\documentstyle[12pt]{article}
\let \ts=\thinspace
\newcommand{\tts}{\kern .08333em}
\newcommand{\negtts}{\kern-.08333em}

\begin{document}
\advance \baselineskip by 10pt
\begin{center}
{\Large\bf THE ASTROPHYSICAL S-FACTOR OF THE REACTION
$^7$Be(p,$\gamma$)$^8$B IN THE DIRECT CAPTURE MODEL}
\end{center}

\begin{center}
{\Large H.~Krauss, K.~Gr\"un, T.~Rauscher, H.~Oberhummer}\\
{\sl Institut f\"ur Kernphysik, TU Wien, Wiedner Hauptstr.~8--10\\
A-1040 Wien, Austria}
\end{center}

\noindent {\bf Abstract.} The astrophysical $S$-factor for the reaction
$^7$Be(p,$\gamma$)$^8$B up to an energy of 2~MeV (c.m.) and the capture
cross section of $^7$Li(n,$\gamma$)$^8$Li up to
1~MeV (c.m.) are calculated
using the Direct Capture model (DC). Both calculations are
in good agreement with experimental data.

\vspace{1cm}

\newpage
\section{Introduction}

The reaction $^7$Be(p,$\gamma$)$^8$B plays an important role in the
so-called ppIII chain in the hydrogen burning of main-sequence stars. A
knowledge of the reaction rate is essential to
determine the branching ratios
between the ppI, ppII and ppIII chains. The magnitude of the
reaction cross section is of special interest
for the solar neutrino problem. The reason for this is that in the
$^{37}$Cl neutrino experiment~\cite{dav89} 77\%, in the Kamiokande II
experiment~\cite{hir89} 100\% and in the gallium
experiments~\cite{ham88,gav88}
11\% of the detected neutrino flux originate from the high-energy
neutrinos emitted in the
ppIII chain~\cite{bah89}.
Therefore, the reaction rate of the above
process determines the high-energy solar neutrino flux.

The reaction considered in this paper has been measured by various
authors at
subCoulomb energies. The most recent data have been obtained in the
energy range $E_{\rm c.m.} = 117-1230\, {\rm keV}$~\cite{fil83}.

The dominance of the direct interaction (DI) mechanism
and the validity of the description
with the potential-model approach below the Coulomb barrier
has been established
in many light-ion reactions (\cite{ohu} and references therein).
In this work we
apply this approach to the above reaction. In the next section
we introduce the direct capture model. In Section 3 the result for
the astro\-physi\-cal $S$-factor is given and compared to experimental data.
Finally, in Section 4 the results are summarized.

\section{Potential Model Approach}

Potential models are based on the description of the
dynamics of the reaction by a Schr\"odinger equation with local optical
potentials in the entrance and/or exit channels.
Such models are the ``Distorted Wave Born Approximation" (DWBA)
\cite{aus70,sat83,gle83} for transfer or the
``Direct Capture" model (DC) \cite{chr61,tom63,rol73}
for capture reactions.

In first order perturbation theory the expression
for the differential cross section of a direct capture reaction
is~\cite{kim87}:
\begin{equation}
{d \sigma_{{\rm DC}} \over d \Omega_\gamma}  =
2 \left( e^2 \over \hbar c \right) \left( m c^2 \over \hbar c \right)
\left( k_\gamma \over k_a \right)^{\!\! 3} {1 \over 2 I_A + 1} \;
{1 \over 2 S_a + 1}
\sum_{M_A M_a M_B \sigma}
\mid T_{M_A M_a M_B, \sigma} \mid^2 \quad .
\end{equation}
Here  $I_A$ ($M_A$) and  $S_a$ ($M_a$) are the spins (their
projections on the z-axis) of target and projectile,
and $\sigma$
is the polarization of the electromagnetic radiation ($\sigma = \pm 1$).
The wave numbers of the emitted $\gamma$-rays and of the
asymptotic relative
wave function in the entrance channel are denoted by $k_\gamma$ and $k_a$,
and $m$ is the reduced mass. We couple the angular momenta in the
spin-orbit representation:
	\begin{eqnarray}
	\vec{l}_a + \vec{S}_a &\hspace{-2mm}=&\hspace{-2mm}
	\vec{j}_a \quad ,\label{lasaja}\\
	\vec{l}_b + \vec{S}_a &\hspace{-2mm}=&\hspace{-2mm}
	\vec{j}_b \quad ,\label{ibsajb}\\
	\vec{I}_A + \vec{I}_B &\hspace{-2mm}=&\hspace{-2mm}
	\vec{j}_b \quad .\label{iajbib}
	\end{eqnarray}
	The coupling and notation are the same as in~\cite{kim87}.

	The transition amplitudes
$T_{M_A M_a M_B, \sigma}$
are expanded in terms of rotation matrix elements
$d_{\delta \sigma}^{\lambda}(\theta)$ with the electromagnetic multipole
$\lambda$ ($\lambda$
= E1, E2, M1, \dots),
\begin{equation}
T_{M_A M_a M_B, \sigma} = \sum_\lambda T_{M_A M_a M_B, \sigma}^\lambda
\, d_{\delta \sigma}^{\lambda}(\theta)\quad ,
\end{equation}
where $\delta = M_A + M_a - M_B$ and $\theta$ is the angle between
${\bf k}_a$ and ${\bf k}_\gamma$.
\newline
The transition amplitudes are proportional to the radial integrals
\begin{equation}
T^\lambda \propto I_{l_b j_b I_B; l_a j_a}^\lambda =
\int dr\: U_{l_b j_b I_B}(r) \:
{\cal O}^\lambda(r) \: \chi_{l_a j_a} (r) \quad , \label{radint}
\end{equation}
where $U_{l_b j_b I_B}(r)$ and $\chi_{l_a j_a} (r)$ are the radial parts
of the bound state wave func\-tion and the distorted wave function
in the entrance channel, respectively.
The functions ${\cal O}^\lambda(r)$ are the radial parts
of the electromagnetic
multipole operators, which were taken in their approximated form,
\begin{equation}
{\cal O}^{\rm M1}(r)\simeq  1,  \quad
{\cal O}^{\rm E1}(r)\simeq  r,  \quad
{\cal O}^{\rm E2}(r)\simeq  r^2 \quad .
\end{equation}

\subsection*{Folding potentials}

We solve the radial integral (\ref{radint}) numerically
using  single folding potentials~\cite{sat79} for the bound state
potential and  the  optical potential,
\begin{eqnarray}
V(r)&=&
\lambda \int d{\bf r}_A \,
\rho_A({\bf r}_A)\,
t(E, \rho_A, |{\bf s}={\bf r}-{\bf r}_A|) \quad ,
\label{pots}
\end{eqnarray}
which were assumed to be real.
Here $r$ is the separation of the centers of mass of the
bound or colliding particles,
$\rho_A$ is the nucleon density of the target and $\lambda$
is a normalization constant.
For the effective nucleon-nucleon
interaction $t$ in the entrance and exit channel we chose
the density dependent
form of the M3Y interaction~\cite{kob84}.

The nucleon density $\rho_A$ is derived from the charge density
distribution $\rho_{\rm p}(r)$ of the target $A$
with the assumption,
that the distribution in the nucleus
is the same for protons and neutrons:
$\rho_{\rm n} = (N/Z)\rho_{\rm p}$~\cite[p.\ts 474]{sat83}.
This leads to
\begin{equation}
\rho_A = \rho_{\rm p} + (N/Z)\: \rho_{\rm p} =
(A/Z)\: \rho_{\rm p}\quad . \label{rhoa}
\end{equation}
The  density $\rho_A(r)$ satisfies
\begin{equation}
4\pi\int \rho_A (r)\, r^2 dr = A,\qquad
\label{rhonorm}
\end{equation}

The normalization constant $\lambda$ of the folding potential
($\lambda_{\rm sc}$
for the potential in the entrance channel and
$\lambda_{\rm b}$ for the bound state potential)
accounts for the interplay of the Pauli principle and
distortion and breakup effects. One of the advantages of the
folding procedure lies in the fact that no open geometrical
parameters exist, since $\lambda_{\rm b}$ is adjusted to reproduce
the experimentally known binding energies and $\lambda_{\rm sc}$
can be adjusted to reproduce scattering phase shifts
(if they are available) or the energies of resonant states (which is
done for $^7$Be(p,$\gamma$)$^8$B). Therefore the form of the optical and
bound state potentials is determined uniquely.

\section{Results}

\subsection*{$^7$Be(p,$\gamma$)$^8$B}

The $^7$Be(p,$\gamma$)$^8$B reaction
is of crucial importance since it leads
to the high-energy $^8$B neutrinos. Unfortunately, there are still
significant experimental uncertainties in the low-energy cross section
for this reaction. The reaction was analyzed
before in the DC model by  using
phenomenological potentials~\cite{tom65,kim83}. Furthermore, this reaction
was also analyzed with the resonating group method (RGM)~\cite{joh90}
and the generator coordinate method (GCM)~\cite{des88,des92}.
The numerical calculations in the DI model in this work
were performed using the direct capture code \mbox{TEDCA}~\cite{haribo}.

\subsubsection*{The folding potential for $^7$Be+p}

Because of the instability of $^7$Be there are no experimental charge
distributions available to compute the nucleon density
with Eq.\ts (\ref{rhoa})~\cite{devries}.
Therefore we chose the same
distribution as that of $^7$Li,
which should give a suitable approximation.
Although
the experimental charge distributions of $^7$Li and $^7$Be are somewhat
different, the resulting potential should be quite similar. We
conclude this from calculations comparing the folding potentials of
$^{11}$B+p and $^{12}$C+p.

The folding potential for $^7$Be+p
was calculated using a harmonic oscillator model for the $^7$Li charge
distribution with parameters
$a=1.77$ fm and $\alpha=0.327$~\cite{devries}.
The depth of the pure folding potential is  43 MeV ($\lambda = 1$),
the volume integral per nucleon is 660 MeV fm$^3$.

\subsubsection*{Calculation of $^7$Be(p,$\gamma$)$^8$B without
spin-orbit coupling}

All contributions to the
transition matrix lead to the ground state of  $^8$B with
spin $J^{\pi}=2^+$.
We use the spectroscopic factors ${\cal S}_{\rm 1p_{3/2}}
= 0.977$ and ${\cal S}_{\rm 1p_{1/2}}=0.0561$ from the shell model
calculations of~\cite{coh67}. The DC cross section is a sum
over the two final state configurations ($j_b = 1/2,\: 3/2$),
\begin{equation}
\sigma_{\rm DC} =
\sum_{l_b\, j_b} C^2 {\cal S}_{l_bj_b}\,
\sigma_{l_b\, j_b} \quad , \label{rolform}
\end{equation}
where $C^2$ is the isospin Clebsch-Gordan coefficient.

A strength factor of
$\lambda_{\rm b}=1.041$ yields the correct separation energy of 138 keV of
the bound proton. The bound state wave function is very diluted due to the
very low binding energy. Therefore we had to calculate radial
contributions to the transition matrix up to 200 fm.

In Fig.~1 the astrophysical $S$-factor is shown.
The $\lambda$-parameter of the scattered wave $\lambda_{\rm sc}=0.969$
was determined with the condition that the resonance
in the p-wave has to be at an energy of 632 keV, which leads to
a resonant M1 transition. This value of the normalization
parameter determines the potential uniquely and
was used for all partial waves in the entrance channel.

In Fig.~2 the radial integrand (6) of the s-wave is plotted
at two different energies
($E_{\rm c.m.}=15$ keV and
$E_{\rm c.m.}=1$ MeV).
At low energies there are contributions to the integral from
very far outside the nucleus. Those contributions can be
neglected only for distances larger than 200 fm.

As one can see in Fig.~1, the  calculated
 $S$-factor of $^7$Be(p,$\gamma$)$^8$B agrees with the
experiments~\cite{fil83,kav69,vau70}
quite nicely in absolute value over a wide range of energies
(100 keV $-$ 2 MeV), except in the resonance energy region near
$E_{\rm c.m.}=632$ keV.
In particular there is also a very good agreement with the
mean experimental value of $S$(0) that
is used by Bahcall~\cite{bah89} in the standard solar model.
A comparison of the parametrization of the potential
model calculation with these experimental
values is given in Tab.~1. The
data of~\cite{kav69} in Fig.~1 were
renormalized as suggested by~\cite{bar86}.

The experimental width of the 1$^+$-resonance at 632 keV is not
reproduced so well. The calculated width
of 100 keV is too large by a factor of 2.5. A possible explanation
for this is the neglect of configuration mixing and
other channels, such as the channel
${\rm p}+^7{\rm Be}_{(1/2)^-}$, which could only be taken into account in a
coupled channel calculation.

\subsubsection*{Calculation with spin-orbit coupling in the entrance
channel}

Currently another resonance is discussed, which should correspond to a
$J^{\pi}=1^+$ state of $^8$B with an excitation energy $E_{\rm x}
\approx$
1.5 MeV. This state is derived from comparison with the excitation spectra
of the mirror nucleus $^8$Li~\cite{ajz84}
($1_1^+:\; E_{\rm x}=0.9808$\ts MeV,
$1_2^+:\; E_{\rm x}=2.255$\ts MeV). Up to now, no experimental
evidence for the $1_2^+$  state in $^8$B is available.
 The $1_2^+$ state is predicted by the GCM~\cite{des88} and also
by the potential model, if one takes spin-orbit coupling into account.
The two states are assumed to have the configurations
$^7$Be+p(1p$_{3/2}$) ($1_1^+$) and $^7$Be+p(1p$_{1/2}$) ($1_2^+$).

We are able to include this  doublet
in our calculation by employing an additional spin-orbit potential in the
entrance channel. For this spin-orbit
potential we used the usual
parametrization~\cite{per76}, with the Woods-Saxon form factor
replaced by the form factor of the
folding potential. The strength $\lambda_{\rm sc}=0.94$, together
with the spin-orbit term ($V_{\rm so}=2.44$\ts MeV),
reproduces the correct energy separation of the two 1$^+$-resonances
at $E_{\rm c.m.}=632$ keV
and $E_{\rm c.m.}=1.4$ MeV. Fig.~3 displays a plot of the
contributions of different partial waves and multipolarities.
The influence of the
$J^{\pi}=1_2^+$-resonance on the $S$-factor is negligible, especially for
thermonuclear energies.

\subsubsection*{Calculation with spin-orbit coupling in both
channels}

According to the shell model calculations of ~\cite{coh67},
the $1_1^+$ level in $^8$B has a better p$_{3/2}$
structure (${\cal S}_{\rm 1p_{3/2}}
= 0.3215$ and ${\cal S}_{\rm 1p_{1/2}}=0.1240$). Nevertheless, we also
want to consider the description of  the ground and first
excited (unbound) state of $^8$B as a p-wave spin-doublet.
 The two states should then have the configurations
$^7$Be+p(1p$_{3/2}$) ($2^+$) and $^7$Be+p(1p$_{1/2}$) ($1_1^+$).

The form factor of the spin-orbit potential stays the same as for
the calculation with spin-orbit coupling in the entrance
channel.
The spin-orbit strength $V_{\rm so}=2.01$\ts MeV
was adjusted to describe the
correct energy splitting of the two corresponding states in $^8$Li.
The so determined spin-orbit potential is able to reproduce with
$\lambda_{\rm b}=\lambda_{\rm sc}=1.01733$ both the binding energy
of the ground state and the resonance in the entrance channel at
0.632 MeV. The potentials for the three assumptions (no spin-orbit
potential, spin-orbit in the entrance channel and spin-orbit in
both channels) are shown in Tab.~2.

Fig.~4 displays the $S$-factor calculated with the additional
spin-orbit potential in both channels. Now the
height of the resonance at 632\ts keV is reduced, compared to the
previous calculations. At this energy, the dominating M1-contribution
 is formed by
the overlap  of the quasi-bound and bound state (Eq.~(6)), which are now
almost orthogonal. This is due to the assumption of a spin-dublet with a
spin-orbit potential. In Fig.~5,
the radial wave function of the bound state (full line)
is shown together
with the scattering wave functions of the p wave
at resonance energy (632\ts keV) for the
case of zero spin-orbit potential (dashed line) and spin-orbit
potential (dashed dotted line). The overlap of scattering wave and
bound state reaches its maximum at 3~fm in both cases, but is much
smaller in its absolute value for the case of spin-orbit potential.

As a consequence, the description of the experimental $S$-factor
by the potential model gets worse in the resonance energy region
(Fig~(4)). The assumption
of a spin-dublet does not lead to a good description of the
data.

\subsection*{$^7$Li(n,$\gamma$)$^8$Li}

We now want to calculate the mirror reaction
$^7$Li(n,$\gamma$)$^8$Li with the assumption of
spin-orbit coupling in the entrance
channel only, because this seems to be the most realistic potential.
 Since the determination of the folding potential
is not depending on isospin, the $^7$Li+n-potential is the same as
the $^7$Be+p-potential of the last section. Therfore the optical
 potential for the entrance channel
can be taken from the analog calculation of
$^7$Be(p,$\gamma$)$^8$B (see also Tab.~2).
Also the shell model spectroscopic factors are the same
 for $^8$Li and $^8$B.

We consider the capture into the ground state ($J^\pi=2^+$) and
the first excited state ($J^\pi=1^+$, $E_{\rm x}=0.9808$\ts MeV) of $^8$Li.
The correct binding energies of these states are reproduced with
the parameters $\lambda_{\rm b}=1.0361$ for the ground state and
$\lambda_{\rm b}=0.9627$ for the first excited state of $^8$Li.

The result of the calculation is shown in Fig.~6. The experimental
data were taken from~\cite{wie89} (filled circles) and~\cite{imh59}.
The open circles and triangles result
from two different normalization procedures. We did not attempt to
reproduce the  resonance at $E_{\rm c.m.}=0.222$\ts MeV, which
corresponds to a $3^+$ state in $^8$Li, since this state has a bad
$^7$Li+n structure~\cite{sto77}.

\section{Summary}

We have shown that the reactions $^7$Be(p,$\gamma$)$^8$B
and $^7$Li(n,$\gamma$)$^8$Li can be described
by a potential model at thermonuclear energies.
The theoretical values of the low-energy
astrophysical $S$-factor for the different
assumptions concerning the single-particle configurations of $^8$B
are in good agreement with the mean
experimental values of~\cite{bah89}. It can be concluded
that the direct interaction mechanism is dominant at energies
well below the Coulomb barrier.

The resonance width and strength of the experimentally known resonance at
$E_{\rm c.m.}=632$ keV are both overestimated by the potential model, if
one uses no spin-orbit potential or a spin-orbit potential in the
entrance channel only. In these cases no special assumptions concerning
the single-particle configurations of the final bound state in $^8$B
are made. If one assumes the configurations
$^7$Be+p(1p$_{3/2}$) ($2^+$) and $^7$Be+p(1p$_{1/2}$) ($1_1^+$) for
this state
and introduces an additional spin-orbit potential to describe the
bound and first unbound state simultaneously,
the DC cross section in the energy
region of the resonance is strongly reduced (Fig.~4) and
underestimates the data, due to the overlap of relatively orthogonal
wavefunctions.
But in the non-resonant energy region the reproduction of the cross section
data is excellent in all cases.

The reaction $^7$Li(n,$\gamma$)$^8$Li was calculated with the potential
in the entrance channel taken from $^7$Be(p,$\gamma$)$^8$B, since this seems
to be the most realistic and therefore favorable potential ansatz.
The agreement with the reaction data is good, as can be seen in Fig.~6.

The progress of applying the potential model combined with the
folding procedure lies mainly
in the fact that no parameter has to be adjusted
to the reaction data. This is neither the case in the
phenomenological fits of Barker et al.~\cite{bar86} nor
Kim et al.~\cite{kim87}, where potential parameters have to be
fitted to the experimental reaction data.

In our model the strengths of the folding potentials
are  adjusted to reproduce the energies of the
bound and quasi-bound states. This procedure is similar to
microscopic models~\cite{des88}, where the correct binding
energies are ensured by adjusting a parameter in the
nucleon-nucleon interaction.

Acknowledgement: We want to thank the
\"Osterreichische Nationalbank (project 3924) and the FWF (project
P7838-TEC) for their support.

\newpage
\pagestyle{empty}
\noindent
{\bf Figure captions:}

\bigskip
\noindent
Fig.~1: Astrophysical $S$-factor for the reaction $^7$Be(p,$\gamma$)$^8$B.
Full curve: potential model calculation (no spin-orbit coupling).
The data points are
from~\cite{fil83} (triangles), \cite{kav69} (circles) and
\cite{vau70} (squares).

\bigskip
\noindent
Fig.~2: $^7$Be(p,$\gamma$)$^8$B:
Radial integrand of Eq.\ts (6) for two different energies
in the entrance channel (s-wave only).

\bigskip
\noindent
Fig.~3: Contributions of different partial waves and multipolarities
to the cross section of $^7$Be(p,$\gamma$)$^8$B.
Here an additional spin-orbit potential in the entrance channel was used
which influences the p-wave contribution near $E_{\rm c.m.}=1.4$ MeV.

\bigskip
\noindent
Fig.~4: Astrophysical $S$-factor of $^7$Be(p,$\gamma$)$^8$B.
Full curve: potential model calculation (spin-orbit coupling in both
channels).

\bigskip
\noindent
Fig.~5: Radial wave functions of the $^7$Be+p bound state
(full line)  and scattering wave functions of the p wave
at resonance energy (0.632~MeV) for zero spin-orbit potential
(dashed line) and spin-orbit
potential in both channels (dashed dotted line).

\bigskip
\noindent
Fig.~6: Total cross section
for the reaction $^7$Li(n,$\gamma$)$^8$Li.
The experimental
data are from~\cite{wie89} (filled circles) and~\cite{imh59}
(open circles and triangles).

\newpage
\noindent
{\bf Table captions:}

\bigskip
\noindent
Tab.~1: Comparison of the results of the potential model
calculation of $^7$Be(p,$\gamma$)$^8$B for the three different assumptions
(no spin-orbit coupling, spin-orbit in the entrance channel and
spin-orbit in both channels)
with the experimental values of the astrophysical
$S$-factor and its derivative as given in~\cite{bah89}.

\medskip
\noindent
Tab.~2: Comparison of the potentials used in the
calculation of $^7$Be(p,$\gamma$)$^8$B and $^7$Li(n,$\gamma$)$^8$Li.

\newpage
\begin{table}
\begin{center}
\begin{tabular}{|c|c|c|c|c|}
\hline
\rule[-8pt]{0pt}{24pt}
 & Exp. [5]  & without SO & SO$_{(i)}$ & SO$_{(i+f)}$
\rule[-2mm]{0cm}{2mm} \\ \hline \hline
\rule{0cm}{16pt}
$S(0)$ & $0.0243\pm 0.0053$ & $0.0249$ &
$0.0249$ & $0.0236$
\rule[-2mm]{0cm}{2mm}\\ \hline
\rule{0cm}{16pt}
$\dot{S}(0)$ & $-3\cdot10^{-5}$ & $-3.2\cdot10^{-5}$ &
$-3.2\cdot10^{-5}$ & $-3.1\cdot10^{-5}$
\rule[-2mm]{0cm}{2mm}\\ \hline
\rule{0cm}{16pt}
$S(18\, {\rm keV})$ & $0.0238\pm0.0052$ &
$0.0243$ & $0.0243$&
$0.0231$
\rule[-2mm]{0cm}{2mm}\\ \hline
\end{tabular}
\end{center}
\end{table}

\newpage
\begin{table}
\begin{center}
\begin{tabular}{|c|c|c|c||c|}
\hline
\rule[-8pt]{0pt}{24pt}
&\multicolumn{3}{c||}{$^7$Be(p,$\gamma$)$^8$B}  & $^7$Li(n,$\gamma$)$^8$Li
\rule[-2mm]{0cm}{2mm}\\ \hline
  \rule[-8pt]{0pt}{24pt} & without SO & SO$_{(i)}$ & SO$_{(i+f)}$ & SO$_{(i)}$
\rule[-2mm]{0cm}{2mm} \\ \hline \hline
\rule{0cm}{16pt}
$\lambda_{\rm b}$ & $1.041046$ & $1.041046$ & $1.01733$
 & $1.0361$ \hspace{1mm} $0.9627$
\rule[-2mm]{0cm}{2mm}\\ 
\rule{0cm}{16pt}
$V_{\rm so}$  & 0 & 0 & $2.01$ & 0
\rule[-2mm]{0cm}{2mm}\\ 
\rule{0cm}{16pt}
$E_{\rm res}$ &
$-0.138$ & $-0.138$&
$\overbrace{\rule{0cm}{14pt} -0.138 \hspace{6mm} \mbox{---}}$ &
$-2.033 \hspace{3mm} -1.052$
\rule[-2mm]{0cm}{2mm}
\rule[-2mm]{0cm}{2mm}\\ 
\rule{0cm}{16pt}
$j_b$ &
 1/2, 3/2 & 1/2, 3/2 &
\hspace{1mm} 3/2 \hspace{1mm} 1/2&
1/2, 3/2 \hspace{1mm} 1/2, 3/2
\rule[-4mm]{0cm}{4mm}\\ \hline
\rule{0cm}{16pt}
$\lambda_{\rm sc}$ & $0.96895$ & $0.94$ & $1.01733$ & $0.94$
\rule[-2mm]{0cm}{2mm}\\ 
\rule{0cm}{16pt}
$V_{\rm so}$  & 0 & $2.44$ & $2.01$ & $2.44$
\rule[-2mm]{0cm}{2mm}\\ 
\rule{0cm}{16pt}
$E_{\rm res}$ &
$0.632$ & $\overbrace{\rule{0cm}{14pt} 0.632\qquad 1.4}$ &
$\overbrace{\rule{0cm}{14pt} \hspace{2mm}\mbox{---} \qquad  0.632}$
& $\overbrace{\rule{0cm}{14pt}
\hspace{1mm}\mbox{---}\hspace{14mm} \mbox{---}}$
\rule[-2mm]{0cm}{2mm}\\ 
\rule{0cm}{16pt}
$j_a$ &
 1/2, 3/2 & \hspace{1mm} 3/2 \hspace{6mm} 1/2 &
3/2 \hspace{5mm} 1/2 &  3/2 \hspace{9mm} 1/2
\rule[-2mm]{0cm}{2mm}\\ \hline
\end{tabular}
\end{center}
\end{table}

\end{document}